\documentclass[12pt]{iopart}

\usepackage{graphicx}% Include figure files
\usepackage{bm}% bold math

\def\abinitio{\textit{ab initio}}
\def\XdS{X$^{2}\Sigma^{+}$}
\def\AdP{A$^{2}\Pi$}
\def\BdS{B$^{2}\Sigma^{+}$}
\def\aqP{a$^{4}\Pi$}
\def\cm{cm$^{-1}$}
\def\bhp{BH$^+$}
\def\alhp{AlH$^+$}
\def\APiXSig{A$^2\Pi$-X$^2\Sigma^+$}
\def\dSp{$^2\Sigma^+$}
\def\dSpg{$^2\Sigma^+_g$}
\def\dS{$^2\Sigma$}
\def\dP{$^2\Pi$}
\def\dD{$^2\Delta$}
\def\re{$r_e$}
\def\we{$\omega_e$}
\def\wexe{$\omega_e x_e$}
\def\A{$\mathcal{A}$}
\def\B{$\mathcal{B}$}

\begin{document}

\title{Challenges of laser-cooling molecular ions}

\author{Jason~H~V~Nguyen$^{1,4}$, C~Ricardo~Viteri$^{2,4,5}$, Edward~G~Hohenstein$^{3}$, C~David~Sherrill$^{3}$, Kenneth~R~Brown$^{2}$ and Brian~Odom$^{1}$}

\address{$^{1}$ Department of Physics and Astronomy, Northwestern University, 2145 Sheridan Road, Evanston, Illinois, USA 60208}
\address{$^{2}$ Schools of Chemistry and Biochemistry; Computational Science and Engineering; and Physics, Georgia Institute of Technology, Atlanta, Georgia 30332, USA}
\address{$^{3}$ Center for Computational Molecular Science and Technology, School of Chemistry and Biochemistry, and School of Computational Science and Engineering, Georgia Institute of Technology, Atlanta, Georgia 30332, USA}
\address{$^{4}$These authors contributed equally.}
\address{$^{5}$Present address: Entanglement Technologies, Inc., 42 Adrian Ct., Burlingame, CA 94010}

\ead{\mailto{ken.brown@chemistry.gatech.edu}, \mailto{b-odom@northwestern.edu} }

\date{\today}

\begin{abstract}
The direct laser cooling of neutral diatomic molecules in molecular beams suggests that trapped molecular ions can also be laser cooled. The long storage time and spatial localization of trapped molecular ions provides the opportunity for multi-step cooling strategies, but also requires a careful consideration of rare molecular transitions. We briefly summarize the requirements that a diatomic molecule must meet for laser cooling, and we identify a few potential molecular ion candidates. We then perform a detailed computational study of the candidates \bhp\ and \alhp, including improved  {\em ab initio} calculations of the electronic state potential energy surfaces and transition rates for rare dissociation events.  Based on an analysis of population dynamics, we determine which transitions must be addressed for laser cooling and compare experimental schemes using continuous-wave and pulsed lasers.
\end{abstract}

\submitto{\NJP}
\maketitle

\section{Introduction}
The long-held notion that laser-cooling molecules is infeasible has been recently overturned by the transverse laser-cooling of SrF \cite{ShumanPRL2009, ShumanNAT2010}. The direct Doppler cooling was possible due to the nearly diagonal Franck-Condon factors (FCFs) of the A-X transition and strong optical forces resulting from the short excited state radiative lifetime. Diagonal FCFs minimize the number of vibrational states populated by spontaneous emission during the cooling time. Similar to previous proposals~\cite{DiRosaEPJD2004, StuhlPRL2008}, the number of relevant rotational states involved in the cooling cycle is minimized by judicious choice of the initial angular momentum state~\cite{ShumanPRL2009}.

In ion traps, laser-cooled atomic ions can be used to sympathetically cool any other co-trapped atomic or molecular ion species~\cite{DrewsenIJMS2003, RyjkovPRA2006, WillitschPCCP2008}, and this technique is being used to study gas-phase atomic and molecular ions at very low temperatures in laboratories around the world. For example, sympathetically cooled ions have been used for mass spectrometry~\cite{BabaJJAP1996, DrewsenPRL2004}, precision spectroscopy~\cite{KoelemeijPRL2007, RosenbandPRL2007, WolfPRA2008V78, HerrmannPRL2009, BatteigerPRA2009, WolfPRL2009}, and reaction measurements~\cite{HojbjerrePRA2008, OffenbergJPB2009, RothPRA2006V73, OkadaJPB2003, StaanumPRL2008, WillitschPRL2008, GingellJCP2010}. Because the impact parameter in ion-ion collisions is much larger than the molecular length scale, the internal states of sympathetically cooled molecular ions are undisturbed.  A low-entropy internal state can be prepared either by producing the molecular ion through state-selective photoionization ~\cite{TongPRL2010} or by taking advantage of the long lifetime of the sympathetically cooled molecular ions to optically pump the internal degrees of freedom~\cite{VogeliusPRA2004, StaanumNatPhys2010, SchneiderNatPhys2010}. Now that internal state control of sympathetically cooled molecular ions has been demonstrated, it is interesting to consider the possibility of eliminating the atomic coolant ions (and the accompanying laser equipment) and directly Doppler cooling certain molecular ions.  Since molecular ions remain trapped for exceedingly long times independent of internal state, ion traps relax requirements on molecular transition moments and repump rates, thus offering a unique environment for molecular laser cooling.  Direct cooling would also be desirable when the coolant ion may create unwanted complications (e.g.,  quantum information processing using trapped polar molecular ions coupled to external circuits~\cite{2009arXiv0903.3552S}), or when the light driving the atomic cycling transition results in a change in the molecules' internal state (e.g., resonant absorption or photodissociation).  Achieving the necessary cycling of a strong transition would also allow straightforward direct fluorescence imaging of trapped molecular ions at the single-ion level. 

On the basis of spectroscopic data available in the literature (see the appendix), our molecular ion survey found \bhp\ \cite{GuestCPL1981, KleinJCP1982, Ramsay1982, KusunokiCPL1984, ViteriJCP2006, ShiIJQC2010} and \alhp\ \cite{GuestCPL1981AlHp, KleinJCP1982, MullerJCP1986, MullerZN1988, LiCPB2008} to be among the most promising candidates. In this manuscript, we review all identified challenges of maintaining a closed excitation scheme for direct laser cooling of \bhp\ and \alhp\ stored in ion traps. Unlike for the case of SrF, where slow vibrational decay relative to the interaction time allows a straightforward probabilistic approach to predict repump requirements \cite{ShumanPRL2009}, designing a laser cooling experiment for trapped ions requires careful modeling of vibrational decays within the ground state, since they result in diffusion of parity and rotational quantum numbers. Additionally, for \bhp\ and \alhp\ as with all molecular Doppler cooling experiments, rare decay and photofragmentation processes need to be considered, since they can occur on timescales comparable to Doppler cooling and much shorter than the ion trapping lifetime.

This paper is organized as follows. In section~\ref{sec:gen-exp-considerations} we discuss the molecular properties required for Doppler cooling.  In section~\ref{sec:laser-cooling}, we consider in detail two cooling candidates, \bhp\ and \alhp, present new quantum chemical calculations for these species (section~\ref{subsec:abinitio}), discuss the calculation of the Einstein coefficients (section~\ref{subsec:EAB_coeff}), and present the rate-equation model used to determine the number of cooling photons scattered (section~\ref{sec:rate-eqns}). In section~\ref{subsec:cooling-scheme}, we discuss the cooling scheme, including the effect of spin-rotation and spin-orbit splitting of the \dSp\ and \dP\ states, respectively.  The results of our simulation are in section~\ref{subsec:sim-results}, and we present our calculations for rare events, which terminate the cycling transition, in section~\ref{subsec:disastrous-events}.  In section~\ref{sec:prospects}, we discuss the technological requirements to carry out the proposed experiments, and we propose the use of femtosecond lasers to provide multiple repump wavelengths from a single source. Finally, section~\ref{sec:conclusions} summarizes the concepts, simulations and prospects for Doppler cooling of molecular ions. In the appendix, we present a few additional classes of molecules that need to be studied in more detail to decide whether they are candidates for direct laser cooling experiments. 

\section{\label{sec:gen-exp-considerations}General experimental considerations}
Doppler laser cooling requires the scattering of many photons, with each emitted photon on average carrying away an amount of energy proportional to the laser detuning from resonance~\cite{WinelandPRA1979}. To cool a two-level particle with a mass of $\sim 10$ amu and a visible transition linewidth of $\sim 1$ MHz from room temperature to millikelvin temperatures requires the scattering of $10^4-10^6$ photons.  Two-level systems are easily obtained in atoms, {\em e.g.},  ${^{2}\mathrm{P}}$-${^{2}\mathrm{S}}$ transitions in alkali metals (for neutrals) and alkaline earth metals (for ions). In both cases, the ground states have a closed shell with one valence electron, and the state of the atom is well described by the orbital and spin angular momentum of the valence electron~\cite{Metcalf_book}.

Electronic two-level systems are impossible to obtain in molecules since the vibrational and rotational degrees of freedom introduce multiple decay paths not present in atoms.  Spontaneous decay into non-cycling bound or dissociative states generally terminates the cycling transition long before the Doppler cooling limit is reached; however, a carefully chosen molecule can significantly reduce the probability of such decays. The ideal molecule would possess an excited and electronic ground state with equivalent potential surfaces separated by an energy offset in the optical region, resulting in perfectly diagonal FCFs. Most intuitive of the types of transitions which might lead to diagonal FCFs \cite{IsaevPRA2010}, are transitions which excite a single electron from the highest occupied molecular orbital (HOMO) to the lowest unoccupied molecular orbital (LUMO). This transition will have negligible effect on the bond length when both the HOMO and LUMO are either non-bonding or anti-bonding orbitals ~\cite{MillerPT2010}. Alternatively, one may choose a molecule with an unpaired electron and a corresponding hole in one of the highest bonding orbitals, in which the optical transition moves the hole to another bonding orbital.

In addition to considering the FCFs, one must take into account the effect of rotational branching due to spontaneous emission, unexpected decay paths due to the break down of the Born-Oppenheimer approximation and dipole-forbidden transitions.

\subsection{\label{subsec:diag-FCF}Diagonal Franck-Condon factors}
When a decay from the excited state to the ground state occurs, the branching ratio into final vibrational states is determined from the Franck-Condon overlap, which measures the similarity between the excited and ground vibrational wavefunctions \cite{Herzberg}.   Molecules which have similar excited-state and ground-state potential energy curves (PECs) reduce the total number of repumping lasers required, so it is useful to determine the repumping requirements as the two PECs are made more dissimilar.

\begin{figure}
\center
\includegraphics[width=0.65\textwidth]{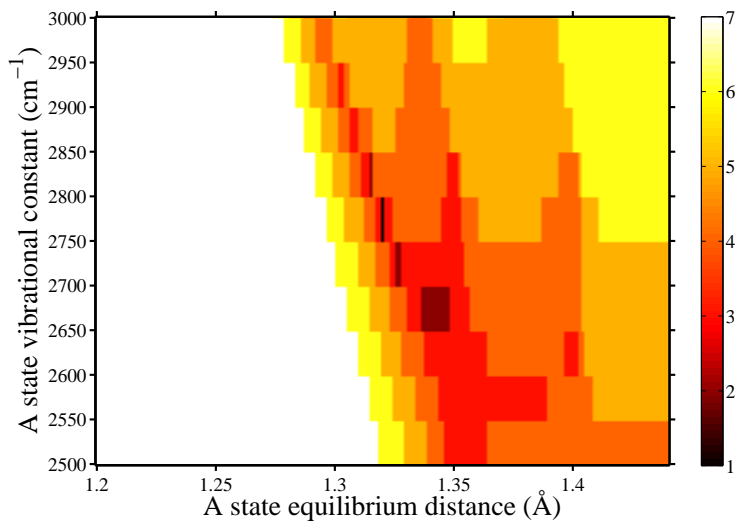}
\caption{\label{fig:franckmorse}The effect of excited state vibrational constant and equilibrium bond length on the number of optically pumped vibrational states required to have less than 1 part-per-million population loss after spontaneous emission. Calculations are based on probabilistic analysis of FCFs of an hypothetical $A-X$ system. White space represents parameter regions in where at least 7 vibrational bands must be addressed. The ground and excited state PECs are represented by Morse functions with a fixed anharmonicity of \wexe\ of 50 \cm. The ground state, $X$, has an equilibrium distance $R_e$ of 1.32 \AA\ and a vibrational constant \we\ of 2750 \cm.  Laser cooling with one laser is possible only in a narrow region (in black) where the $A$-state spectroscopic constants are similar enough to $R_e$ and \we.}
\end{figure}

High-accuracy quantum-chemical calculations of the excited PEC are not typically available; however, the Morse potential, with only three parameters, is a reasonable approximation for many bound electronic states.  This approximation is particularly true near the bottom of a potential well, which is energetically distant from the region of non-adiabatic couplings that alter the dissociative asymptotes of the potentials. Assuming excited vibrational levels have infinite lifetimes, the number of transitions that are required to approximately close a cycling transition depends strongly on all of the three parameters. Figure \ref{fig:franckmorse} shows the effect of changing the equilibrium bond distance and vibrational frequency for fixed anharmonicity. The number of vibrational levels that must be addressed to have less than 1 part-per-million population loss is very sensitive to differences in the ground and excited state bond lengths. In the ideal case of perfectly diagonal FCFs, a single laser is required; a difference of only 5\% in the bond lengths requires from four to seven lasers.

\subsection{\label{subsec:rotational}Elimination of rotational branching}
In the dipole approximation, each emitted photon can change the total angular momentum of the molecule by 0, or $\pm 1\ \hbar$. The emission results in the population of additional states through Stokes and anti-Stokes processes. Each additionally populated angular momentum state will need to be addressed in order to achieve a high scattering rate and efficient laser cooling.

The number of angular momentum states populated depends on which Hund's case (a) or (b) is important~\cite{Herzberg}. For the purpose of our discussion here, we will initially ignore spin (equivalent to Hund's case (b)) and later look at complications which occur in Hund's case (a). We will also ignore nuclear spin and the resulting additional hyperfine states and transitions (the resultant hyperfine splitting is sufficiently small that in practice these additional states can easily be addressed by use of an electro-optical modulator (EOM)\cite{ShumanPRL2009,ShumanNAT2010}).

\begin{figure}
\center
\includegraphics[width=0.65\textwidth]{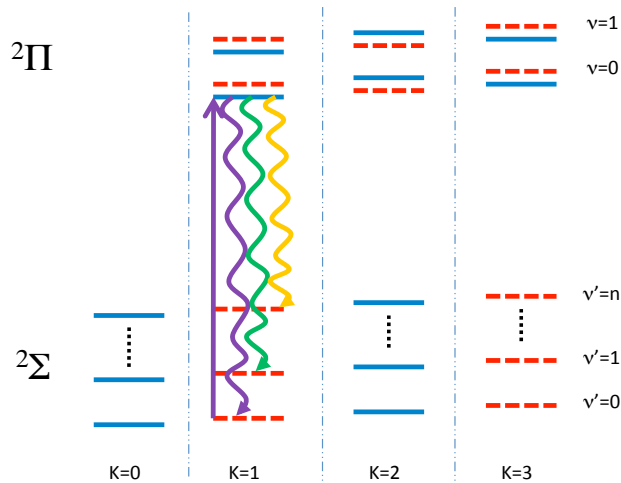}
\caption{\label{fig:leveldiagram}Energy level schematic of proposed $^2\Pi$-$^2\Sigma$ transition. Red dashed levels are odd parity and blue solid levels are even parity. $K$ refers to the total angular momentum without spin, and $v$ and $v'$ label the vibrational levels of the excited and ground states. Starting in the $K=1$ state and exciting the $Q$-branch (solid arrows) results in a closed transition modulo the vibrational states (wiggly arrows).}
\end{figure}

$\Lambda-$doubling breaks rotational symmetry about the bond-axis, resulting in electronic states of well-defined symmetry relative to the plane defined by the axis of nuclear rotation and the bond-axis. A key feature of the SrF experiment \cite{ShumanPRL2009, ShumanNAT2010} is pumping from the ground state $v^\prime$=0, N$^\prime$=1, K$^\prime$=1 state with negative parity to the excited $v$=0, N=0, K=1 state with positive parity, where N is the rotational angular momentum of the molecule, and K is the total angular momentum without spin. As a result, decay from the excited state is limited to those $v$ allowed by Franck-Condon factors and only via the $Q$ branch. The desired transitions and the expected spontaneous emission channels are shown in figure \ref{fig:leveldiagram}.

This method requires the control of the initial angular momentum population, as not all the molecules in the experiment start out with the desired $N$ or $J$.  However, the preparation of an specific initial state of the molecular ion can be achieved by laser-cooling the internal states~\cite{VogeliusPRA2004, StaanumNatPhys2010, SchneiderNatPhys2010}, by cooling the internal states through interaction with cold neutral atoms~\cite{HudsonPRA2009}, using a cryogenic ion trap~\cite{LabaziewiczPRL2008} to limit the number of energetically accessible states and the rate at which they mix, or using state-selective photoionization of a neutral molecule~\cite{TongPRL2010}.

Additional rotational states can be populated by blackbody redistribution between different rotational states, or by the decay of vibrationally excited states through spontaneous emission.  This process does not occur for homonuclear diatomic ions. For heteronuclear diatomic ions, blackbody redistribution results in an increase in the number of required repumps if the redistribution occurs on cooling timescales.  For \bhp\ and \alhp\ we find that rotational redistribution requires tens of seconds, which is longer than the cooling time, and is included in our simulation (see section \ref{subsec:sim-results}).  Vibrational radiative relaxation can limit the number of vibrational bands that must be addressed, at the cost of repumping from a larger set of rotational states due to the diffusion in $J$ during the vibrational cascade.

If vibrational relaxation is sufficiently fast, a probabilistic FCF approach to determine repumping requirements is no longer valid, since the FCF approach assumes vibrational levels are infinitely long-lived.  For example, if we cycle on the $v'=0\leftrightarrow v=0$ transition, and repump on the $v'=1\leftrightarrow v=0$ transition, the ion will occasionally decay via the $v=0\rightarrow v'=2 \rightarrow v'=1$ and  the $v='1 \rightarrow v'=0$ channels, resulting in parity flips and occupation of higher rotational states.

A simple figure of merit (FOM) for the probabilistic FCF approach is calculated by comparing the total time required to Doppler cool the molecule to the vibrational decay rate, since all levels are approximately equally populated at steady-state. The FOM is given by $\frac{A_{00'}}{N A_{1'0'}}$, in which $A_{00'}$ is the excited-state decay rate, $N$ is the total number of photons scattered to Doppler cool, and $A_{1'0'}$ is the vibrational decay rate.  Using an estimate of $N\approx 10^{5}$,  we calculate a FOM of $1$, $20$, $250$ for \bhp, \alhp, and SrF, respectively. We expect that the effect of vibrational decay is most significant for \bhp, followed by \alhp, which have much smaller FOMs than does SrF.  Our rate-equation simulations (see section \ref{subsec:sim-results}) confirm this prediction, with vibrational decay in \bhp\ and \alhp becoming important at short and long times, respectively, and vibrational decay in SrF unimportant on cooling timescales.

\subsection{\label{sec:disastrous-events}Other decay paths}
The discussion so far has relied on the Born-Oppenheimer approximation and only considered dipole-allowed electronic transitions. These approximations reliably describe the molecules of interest for timescales up to a few microseconds. However, long ion-trap lifetimes require the consideration of slow processes driven by violations of the model.

Ideally, the A state is stable against dissociation.  However, constant excitation of the molecule allows for rare events to occur, such as dissociation via the repulsive part of the ground state potential.  This dissociation can be caused by spin-orbit coupling or $L$-uncoupling, which violates the Born-Oppenheimer approximation by mixing the ground and excited electronic state potentials~\cite{Field's_book}. In the case of $L$-uncoupling, it is possible to pick excited states such that symmetry effects forbid dissociation from occurring. Spin-orbit coupling also exists, but the calculated dissociation rate ranges from kHz to Hz and, to our knowledge, has never been measured. (In most molecular beam experiments these rare events are unobserved, since the dissociation timescales are long compared to laser-interaction timescales.) It is also possible that cooling or repumping lasers can cause photodissociation by coupling to the A state to a higher-lying dissociative state.

Additionally, transitions that are not electric dipole allowed can occur due to mixing of parity states by electric fields. For a compensated ion trap, the stray DC electric field at the axis can be effectively cancelled \cite{Berkeland1998}.  Population of opposite parity states is inevitable in the long-time limit, due to magnetic dipole transitions (those of $\Delta J = 0, \pm 1$ terminate on states of the wrong parity). These transitions occur at a rate of 1 transition or less in 10$^5$, also requiring attention when forming a highly-closed cooling cycle~\cite{DiRosaEPJD2004}.

\begin{figure}
\center
\includegraphics[width=0.65\textwidth]{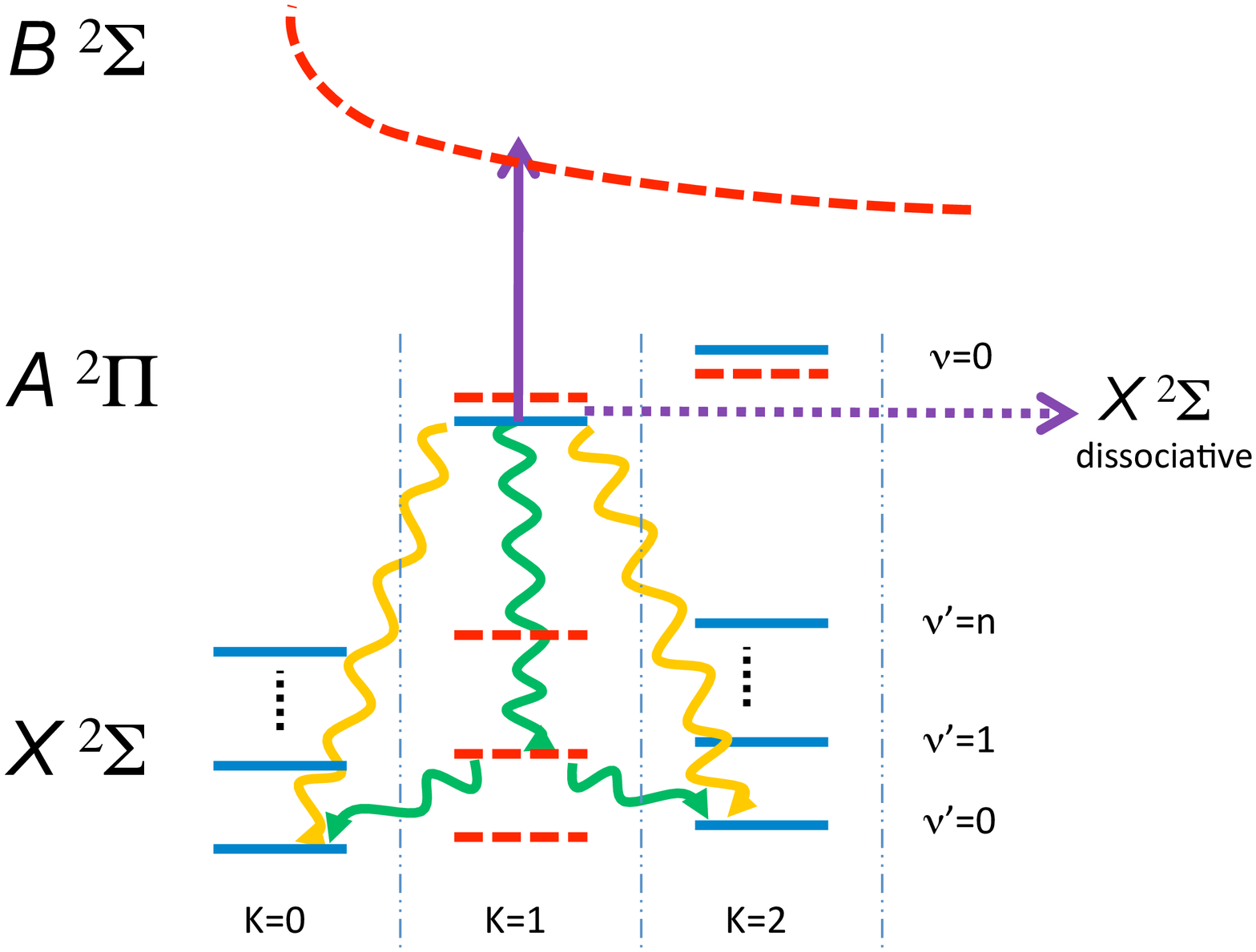}
\caption{\label{fig:badtransitions}Energy level schematic of proposed $^2\Pi$-$^2\Sigma$ transition showing transitions that result in either dissociation or occupation of a rotational state outside of the closed transition.}
\end{figure}

Figure \ref{fig:badtransitions} graphically summarizes all the mechanisms by which the diatomic molecular ion can leave the cycling transition, including optical photon emission to an excited vibrational state followed by the emission of an infrared photon that transfers population to a lower vibrational state. As mentioned in the previous subsection, while this process leads to diffusion in the rotational states and introduces states of additional parity, it may ease the experimental requirements by reducing the number of vibrational bands that must be addressed.

\section{\label{sec:laser-cooling}Example laser cooling cycle for \bhp\ and \alhp}
The \APiXSig\ system of both \bhp\ and \alhp\ are well studied in emission from hollow cathode discharges~\cite{AlmyPR1934, AlmyPR1937, Ramsay1982}, and chemiluminescent ion-molecule reactions~\cite{OttingerJCP1981,MullerJCP1986} have yielded detailed rotational constants for the first two vibrational levels of the X$^2\Sigma^+$ ground state along with low-resolution spectra of a few diagonal vibrational bands. For the BH$^+$ cation, 15 rovibrational energy levels of the ground state (v=0-4 and N=0-2) have been assigned by the extrapolation of photo-selected high Rydberg series of neutral BH~\cite{ViteriJCP2006}. During those experiments, a very strong optical excitation of the ion core was observed when the energy of the scanning laser matched the A-X transition of the cation~\cite{ViteriPRA2007}. Photon absorption rates observed during this isolated core excitation were consistent with Einstein transition probability coefficients calculated theoretically by Klein \textit{et al.}~\cite{KleinJCP1982}. Another favorable property of this molecular ion is that the first quartet states are energetically above the A-X system~\cite{KusunokiCPL1984, HirataCPL2000}, eliminating the possibility of intervening electronic states of different spin to which the upper state could radiate and terminate the cycling transition. With all this information, it is possible to consider a laser cooling experiment for trapped BH$^+$ ions. Similarly, we expect \alhp\ to share most of these favorable characteristics.

\subsection{\label{subsec:methods}Methods}
\subsubsection{\label{subsec:abinitio}Ab initio calculations for \bhp\ and \alhp}
To study the processes outlined in section \ref{sec:gen-exp-considerations}, potential energy functions and dipole moments for the \XdS, \AdP, \BdS, and \aqP\ states of \bhp\ and \alhp\ are required. Additionally, the transition dipole moments and spin-orbit coupling matrix elements between these states are needed. For \bhp, it is possible to solve the electronic Schr\"{o}dinger equation exactly (within a basis) through full configuration interaction (FCI). By constraining the 1s orbital of B to be doubly occupied, a large aug-cc-pV5Z basis \cite{KendallJCP1992} can be used to obtain the potential energy functions. The spectroscopic constants obtained at this level of theory [FCI(3e$^-$)/aug-cc-pV5Z] accurately reproduce the experimentally
determined values (shown in table \ref{tab:spec-cons}), and were computed by fitting a fourth-order polynomial to five energy points centered at R$_e$ and evenly spaced by 0.005 \AA. For \alhp, FCI was not applicable because correlating the important 2s and 2p electrons of Al leads to too
many determinants in the expansion of the wavefunction.  Coupled-cluster methods and their equation-of-motion variants provide a practical alternative~\cite{PurvisJCP1982,StantonJCP1993}. In particular, CC3 and EOM-CC3 \cite{KochCPL1995} were applied with an aug-cc-pCVQZ basis set~\cite{KendallJCP1992}. Again, the 1s orbital was constrained to be doubly occupied. The FCI and EOM-CC3 computations were performed with PSI 3.4~\cite{CrawfordJCC2007}.

Dipole moments, transition dipole moments, and spin-orbit coupling matrix elements were obtained from multireference CI wavefunctions (MRCI). The
MRCI wavefunctions add single and double excitations to a state-averaged complete active space self-consistent field (SA-CASSCF) reference \cite{RuedenbergIJQC1979,RoosCP1980}. The SA-CASSCF orbitals were optimized with an active space consisting of three electrons in the
$2\sigma1\pi_x1\pi_y3\sigma4\sigma5\sigma$ and $4\sigma2\pi_x2\pi_y5\sigma6\sigma7\sigma$ orbitals for \bhp\ and \alhp, respectively. In the MRCI wavefunctions, single and double excitations from the 1s B orbital and from the 2s and 2p Al orbitals were allowed. Again, the aug-cc-pV5Z and aug-cc-pCVQZ basis sets were used for \bhp\ and \alhp, respectively. The MRCI computations were performed with MOLPRO~\cite{MOLPRO}.

All of the PECs, dipole moments, transition dipole moments and spin-orbit matrix elements (doublet-doublet, quartet-doublet and quartet-quartet) generated by the calculations described in this section have been included as a supplementary file.

\begin{table}
\caption{\label{tab:spec-cons}Spectroscopic constants for the \XdS\ and \AdP\ states of \bhp\ and \alhp. Comparisons to previous theoretical and the most precise experimental values, where available, are also given. R$_e$ is the equilibrium bond length, $B_e$, $D_e$, and $\alpha_e$ parameterize the rotational energy level spacings \cite{Herzberg}, \we\ and \wexe\ are the harmonic and anharmonic frequencies of the potential at equilibrium, and D$_0$ is the dissociation energy.  Units are \cm\ unless otherwise specified.}

\tiny
\begin{tabular}{@{}llllllllll}

State & Method & Reference & R$_e$($\AA$) & B$_e$ & D$_e$ & $\alpha_e$ & \we & \wexe & D$_0$(eV) \\
\mr
\bhp \\
\XdS  & MRDCI & \cite{GuestCPL1981} & 1.2059 & 12.552 &  & 0.542 & 2515.2 & 85.86 & 1.86 \\
 & MC-SCF & \cite{KleinJCP1982} & 1.21 & 12.48 &  & 0.475 & 2492 & 64 & 1.78 \\
 & MC-SCF + CI & \cite{KusunokiCPL1984} & 1.208 & 12.53 &  & 0.393 & 2594.8 & 74.94 & 1.92 \\
 & MC-SCF & \cite{BiskupicJMST1988} & 1.2039 & 12.57 & 0.00135 & 0.426 & 2422.7 & 74.61 \\
 & CCSDT(FC)-CBS/mix & \cite{FellerJCP2000} & 1.2047 & 12.58 &  &  & 2524.7 & 64.4 & 1.99 \\
 & MRDCI & \cite{PetsalakisMP2006} & 1.204 & 12.59 &  &  & 2548 & 74.8 & 1.97 \\
 & MRCISD/aug-cc-pV5Z &  & 1.20498 & 12.574 & 0.00125 & 0.47 & 2518.4 & 64.7 & \\
 & FCI(3e$^-$)/aug-cc-pV5Z &  & 1.20484 & 12.578 & 0.00125 & 0.47 & 2519.4 & 64.6 & 1.99 \\
 & Experiment & \cite{Ramsay1982} & 1.20292 & 12.6177 & 0.001225 & 0.4928 & 2526.8$^{\rm{a}}$ & 61.98$^{\rm{a}}$ & 1.95 $\pm$ 0.09$^{\rm{b}}$ \\
 \\
\AdP & MRDCI & \cite{GuestCPL1981} & 1.2158 & 11.649 &  & 0.501 & 2228.3 & 66.88 & 3.33 \\
 & MC-SCF & \cite{KleinJCP1982} & 1.253 & 11.62 &  & 0.467 & 2212 & 52 & 3.2 \\
 & MC-SCF + CI & \cite{KusunokiCPL1984} & 1.247 & 11.76 &  & 0.41 & 2351.8 & 71.38 & 3.19 \\
 & MRDCI & \cite{PetsalakisMP2006} & 1.247 & 11.75 &  &  & 2257 & 52 & 3.3 \\
 & MRCISD/aug-cc-pV5Z &  & 1.24770 & 11.728 & 0.00128 & 0.46 & 2245.0 & 53.4 & \\
 & FCI(3e$^-$)/aug-cc-pV5Z &  & 1.24648 & 11.751 & 0.00128 & 0.46 & 2251.6 & 53.6 & 3.35 \\
 & Experiment & \cite{Ramsay1982} & 1.24397 & 11.7987 &  & 0.4543 \\
 \\
\BdS & MRDCI & \cite{GuestCPL1981} & 1.9036 & 5.037 &  & 0.075 & 1263 & 33 & 1.28 \\
 & SDT-CI & \cite{GuestCPL1981} & 1.9031 & 5.04 &  & 0.074 & 1258 & 31 & 1.28 \\
 & MC-SCF & \cite{KleinJCP1982} & 1.912 & 4.99 &  & 0.097 & 1235 & 32 & 1.26 \\
 & MC-SCF + CI & \cite{KusunokiCPL1984} & 1.91 & 5.01 &  & 0.073 & 1206 & 19 & 1.24 \\
 & MRDCI & \cite{PetsalakisMP2006} & 1.889 & 5.12 &  &  & 1285 & 32 & 1.31 \\
 & MRCISD/aug-cc-pV5Z &  & 1.90180 & 5.048 & 0.000322 & 0.074 & 1264.4 & 30.5 & \\
 & FCI(3e$^-$)/aug-cc-pV5Z &  & 1.90116 & 5.051 & 0.000323 & 0.075 & 1264.3 & 29.9 & 1.35 \\
 \\
\alhp \\
\XdS & MRDCI & \cite{GuestCPL1981AlHp} & 1.6098 & 6.698 &  & 0.318 & 1684 & 81 & 0.666 \\
 & MC-SCF & \cite{KleinJCP1982} & 1.608 & 6.71 &  & 0.317 & 1680 & 71 & 0.74 \\
 & MCQDPT & \cite{LiCPB2008} & 1.6 & 6.78 &  & 0.3945 & 1600 & 82 & 0.92 \\
 & MRCISD/aug-cc-pCVQZ &  & 1.60802 & 6.711 & 0.000422 & 0.30 & 1692.4 & 66.2 \\
 & CC3/aug-cc-pCVQZ    &  & 1.60543 & 6.732 & 0.000434 & 0.31 & 1677.0 & 69.1 & 0.73\\
 & Experiment & \cite{MullerZN1988} & 1.605 & 6.736 & 0.0004469 & 0.382 & 1654 & 74 \\
 \\
\AdP & MRDCI & \cite{GuestCPL1981AlHp} & 1.6047 & 6.741 &  & 0.248 & 1727 & 54 & 1.75 \\
 & MC-SCF & \cite{KleinJCP1982} & 1.609 & 6.7 &  & 0.251 & 1683 & 42 & 1.78 \\
 & MCQDPT\footnote{values are for the \dP$_{1/2}$/\dP$_{3/2}$, respectively} & \cite{LiCPB2008} & 1.6 & 6.7779 &  & 0.2668/0.2489 & 1703/1743 & 47/45 & 1.86/1.82 \\
 & MRCISD/aug-cc-pCVQZ  &  & 1.60375 & 6.746 & 0.000412 & 0.25 & 1726.3 & 33.2 & \\
 & EOM-CC3/aug-cc-pCVQZ &  & 1.59126 & 6.853 & 0.000416 & 0.23 & 1759.3 & 42.8 & 1.90 \\
 & Experiment & \cite{MullerZN1988} & 1.595 & 6.817 & 0.0004152 & 0.243 & 1747 & 43 \\
 \\
\BdS & MRDCI & \cite{GuestCPL1981AlHp} & 2.0582 & 4.097 &  & 0.046 & 1322 & 19 & 1.4 \\
 & MC-SCF & \cite{KleinJCP1982} & 2.063 & 4.08 &  & 0.043 & 1326 & 21 & 1.4 \\
 & MRCISD/aug-cc-pCVQZ  &  & 2.05914 & 4.092 & 0.000155 & 0.039 & 1330.5 & 21 \\
 & EOM-CC3/aug-cc-pCVQZ &  & 2.03038 & 4.209 & 0.000158 & 0.039 & 1375.9 & 23 & 1.43 \\

\end{tabular}

\noindent $^{\rm{a}}$ \cite{ViteriJCP2006}, an $\omega_ey_e$ of $\approx -2$ \cm\ has been determined experimentally \\
\noindent $^{\rm{b}}$ \cite{RosmusJCP1977}

\end{table}

\begin{figure}
\centering
\includegraphics[width=0.65\textwidth]{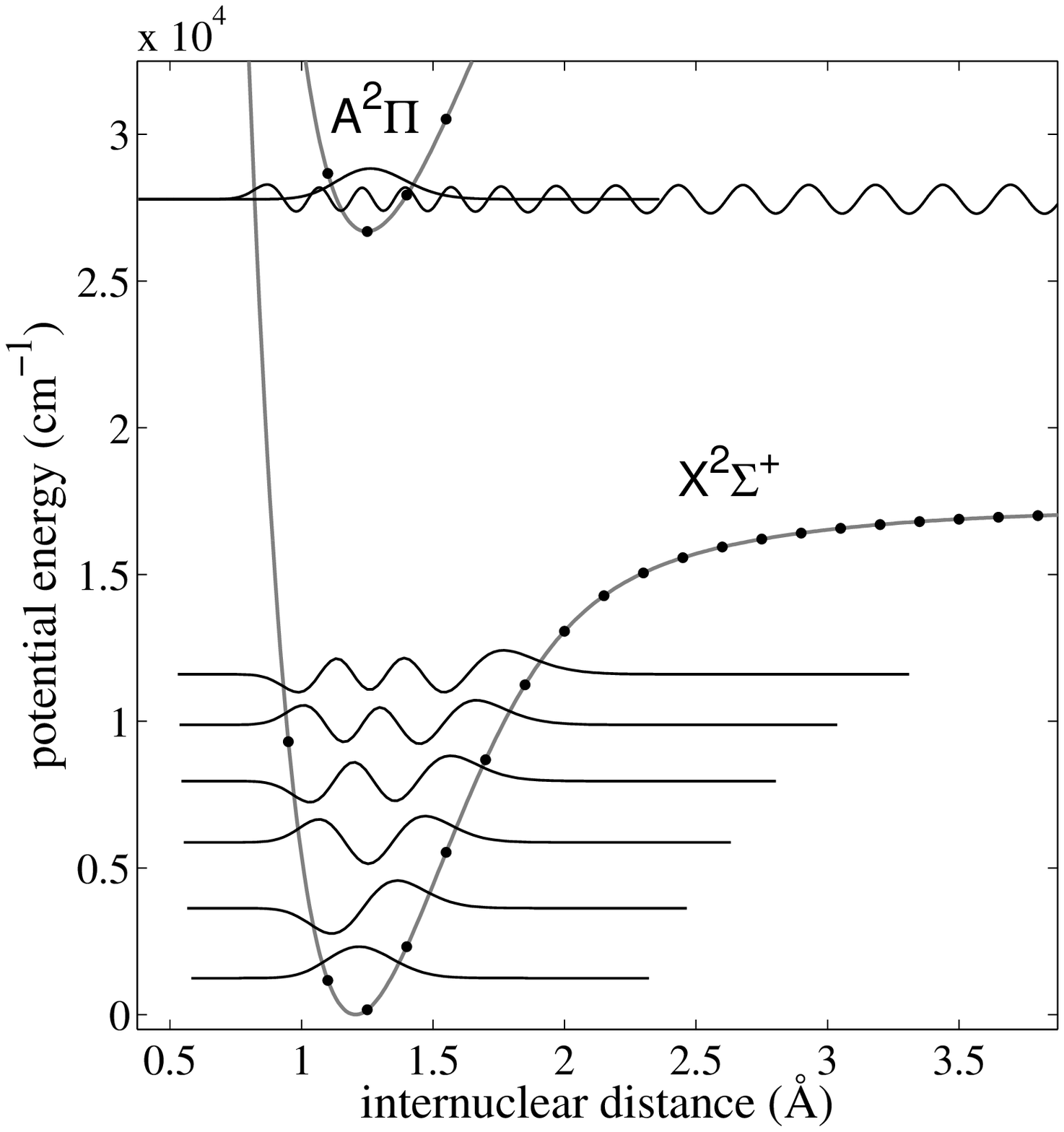}
\caption{\label{fig:fciBH} Born-Oppenheimer potential energy functions of the first two doublet electronic states of \bhp\ at the FCI(3e$^-$)/aug-cc-pV5Z level. Black dots are the actual FCI calculation results, and the solid lines are splines with analytical functions as described in the LEVEL 8.0 and BCONT 2.2 manuals~\cite{bcont2.2,level8.0}. Exact numerical solution of the radial Schr\"odinger equation yields the plotted wavefunctions, which are used to calculate the various decay and emission rates described in the text.}
\end{figure}

\subsubsection{\label{subsec:EAB_coeff}Einstein \A\ and \B\ coefficients}
The Einstein \A\ coefficients connecting two states are calculated using

\begin{equation}
	\mathcal{A}_{u,l}=\frac{\omega_{u,l}^{3}}{3\hbar \pi c^{3} \epsilon_{0}}\frac{g_{u,l}}{2J_{u}+1}\mathcal{S}_{u,l}\mathcal{D}_{u,l}^{2},
\label{eq:EAcoeff}
\end{equation}

\noindent where $\omega_{u,l}/2\pi$ is the transition frequency, $c$ is the speed of light, $\epsilon_{0}$ is the permittivity of free space, $g_{u,l}$ is a degeneracy factor, $J_{u}$ is the total angular momentum of the upper state, $\mathcal{S}_{u,l}$ is the H\"{o}nl-London factor, and $\mathcal{D}_{u,l}$ is the transition dipole-moment matrix element.  The \B\ coefficient is calculated from the \A\ coefficient using

\begin{equation}
	\mathcal{B}_{u,l}=\frac{\pi^{2} c^{3}}{\hbar \omega_{u,l}^{3}}\mathcal{A}_{u,l}.
\label{eq:EBcoeff}
\end{equation}

\subsubsection{Rate-equation model}
\label{sec:rate-eqns}
A rate-equation approach was used to model the population dynamics of \bhp\ and \alhp\, allowing determination of the average number of Doppler-cooling photons scattered before a molecule was pumped into a dark state (ignoring predissociation and photodissociation described in section~\ref{subsec:disastrous-events}). A similar approach has been used for internal-state cooling \cite{VogeliusPRA2004,SchneiderNatPhys2010,StaanumNatPhys2010}. In particular, we solved

\begin{equation}
	\label{eq:rate-eqn-1}
	\frac{d \mathbf{P} }{dt}=\mathbf{M} \mathbf{P},
\end{equation}

\noindent where $P$ is a vector consisting of the $N$ rovibrational levels we include in our model, ordered by increasing energy, and M is a $N \times N$ vector consisting of the Einstein \A\ and \B\ coefficients which couple the different rovibrational states.  Explicitly, the population of a given state follows

\begin{eqnarray}
	\label{eq:rate-eqn-2}
	\fl \frac{dP_{i}}{dt} =  -\sum_{j=1}^{j=i-1}{\mathcal{A}_{ij}P_{i}} - \sum_{j=1}^{j=i-1}{\mathcal{B}_{ij}\rho(\omega_{ij})P_{i}} -\sum_{j=i+1}^{j=N}{\mathcal{B}_{ij}\rho(\omega_{ij})P_{i}} \nonumber \\ +\sum_{j=i+1}^{j=N}{\mathcal{A}_{ji}P_{j}} +\sum_{j=1}^{j=i-1}{\mathcal{B}_{ji}\rho(\omega_{ji})P_{j}}+\sum_{j=i+1}^{j=N}{\mathcal{B}_{ji}\rho(\omega_{ji})P_{j}}. 
\end{eqnarray}

\noindent Here, the $i^{th}$ and $j^{th}$ states are connected by Einstein coefficients denoted by \A$_{ij}$, \B$_{ij}$, and \B$_{ji}$ which correspond to spontaneous emission, stimulated emission, and absorption, respectively, and $\rho(\omega_{ij})$ is the spectral energy density at a given frequency, $\omega_{ij}$.  The Einstein coefficients were calculated using the potential energy curves and dipole moments discussed in section~\ref{subsec:abinitio}.

The average number of Doppler cooling scattered photons was calculated by numerically solving equation (\ref{eq:rate-eqn-1}), and multiplying the population in the excited state of the cycling transition by its spontaneous emission rate. Counting photons emitted into the ground state from the excited state, regardless of whether the excited state was populated by a cycling-laser photon or a repumping-laser photon, is equivalent to  counting the number of cooling-cycle absorption events. A simple estimate of the number of scattering events required for cooling can be obtained by considering the average energy lost per scattering event.  For a laser detuning $\Gamma/2$ from resonance, where $\Gamma$ is the natural linewidth, then the average energy lost per scatter is ${\Delta E = \hbar\Gamma /2}$. The \APiXSig\ linewidths of \bhp\ and \alhp\ are $\Gamma = 2\pi \times 0.7\ \mathrm{MHz}$ and $\Gamma = 2\pi \times 2.6 \ \mathrm{MHz}$, respectively. Starting from $T=300\ \mathrm{K}$, cooling to $\mathrm{mK}$ requires $N\approx 10^{6}$ scattering events for both \bhp\ and \alhp\ for our laser detuning.  We may also take advantage of the Doppler width at higher temperatures by detuning the laser further from resonance to increase the energy lost per scattering event, reducing the laser detuning as the ions are cooled.  Using this method, we estimate a lower limit of $N\approx10^{4}$ scattering events are required to cool to $\mathrm{mK}$.

Finally, it should be noted that a rate-equation approach does not account for coherence effects which can result in dark states.  For example, in atomic systems with a low-lying $D$ state, two lasers are typically used for Doppler cooling to address an $S\rightarrow P$ (cooling) transition and a $D\rightarrow P$ (repumping) transition.  When both lasers are detuned from resonance by similar amounts, dark resonances occur and cycling terminates~\cite{Janik:85}.  The cooling and repumping transitions in our molecular system also exhibit a similar $\Lambda$-system (see figures \ref{fig:coolingoneparity} and \ref{fig:coolingtwoparity}).  Experimentally, these dark resonances can be avoided either by explicitly varying the laser detuning, or by shifting the energy levels by different amounts with an externally applied time-varying field \cite{Janik:85, PhysRevA.65.033413}.

\subsection{\label{subsec:cooling-scheme}Cooling scheme}
\begin{figure}
\center
\includegraphics[width=0.65\textwidth]{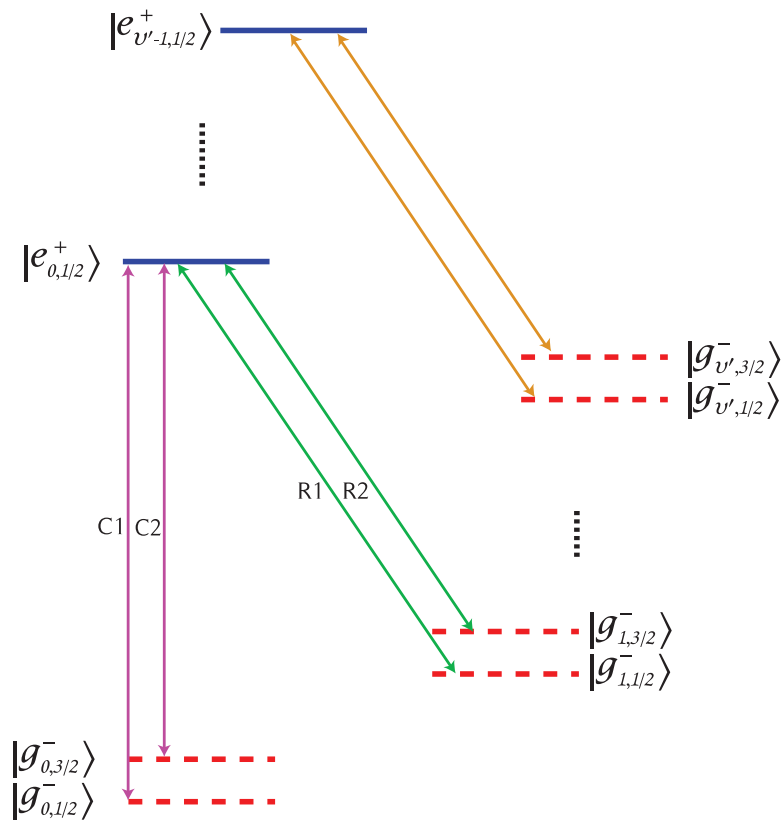}
\caption{Energy level schematic showing cooling and repumping transitions for the case of negligible vibrational and rotational relaxation within the excited and ground state manifolds (not to scale).  Ground and excited states are labeled $|g^{P'}_{v',J'}\rangle$ and $|e^{P}_{v,J}\rangle$, respectively.    \label{fig:coolingoneparity}}
\end{figure}

\begin{figure}
\center
\includegraphics[width=0.65\textwidth]{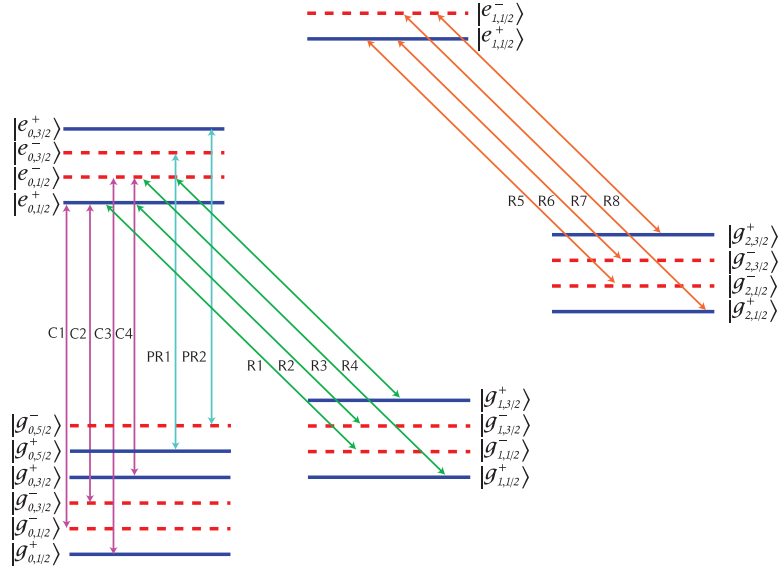}
\caption{Energy level schematic showing cooling and repumping transitions for the case of non-negligible vibrational and rotational transitions within the excited and ground state manifolds (not to scale).  \label{fig:coolingtwoparity}}
\end{figure}

The general cooling scheme, ignoring spin, was discussed in section \ref{sec:gen-exp-considerations} and figure \ref{fig:leveldiagram}.  The inclusion of spin results in spin-rotation splitting for the $^{2}\Sigma^{+}$ states and spin-orbit splitting of the $^{2}\Pi$ states \cite{Herzberg}.  We label the X$^{2}\Sigma^{+}$ and A$^{2}\Pi_{1/2}$ states as $|g^{P'}_{v^{\prime},J'}\rangle$ and $|e^{P}_{v,J}\rangle$, respectively.  The Zeeman substructure is explicitly ignored in our model; however, the resulting reduction in the overall scattering rate due to differing multiplicities of the ground and excited states is included as a degeneracy factor in the Einstein $\mathcal{A}$ and $\mathcal{B}$ coefficients.  In practice, an externally applied magnetic field of a few Gauss is sufficient to lift the degeneracy, and the laser polarization can be chosen to ensure that all Zeeman sublevels are addressed \cite{Janik:85, PhysRevA.65.033413}.   Additionally, both \alhp\ and \bhp\ have hyperfine structure, which is ignored in our model.

We assume that population begins in the $|g^{-}_{0,1/2}\rangle$ state and a laser drives transitions to the $|e^{+}_{0,1/2}\rangle$ state.  From the excited state, population may decay into either the $|g^{-}_{v^{\prime},1/2}\rangle$ state or the $|g^{-}_{v^{\prime},3/2}\rangle$ state, with decreasing probability for increasing $v'$.  The cooling transition, corresponding to $v'=0$, requires driving both the $|g^{-}_{0,1/2}\rangle \longleftrightarrow |e^{+}_{0,1/2}\rangle$ and $|g^{-}_{0,3/2}\rangle \longleftrightarrow |e^{+}_{0,1/2}\rangle$ transitions, as shown in figure \ref{fig:coolingoneparity} with labels C$1$ and C$2$. The ground states belong to the same rotational level ($K'=1$), so this cooling scheme is equivalent to the one discussed in section~\ref{subsec:rotational}.

If vibrational levels with $v^{\prime}\geq 1$ are long-lived, then decay between vibrational levels, which result in parity flips, can be ignored.  In this case, we drive $|g^{-}_{v^{\prime},1/2}\rangle \longleftrightarrow |e^{+}_{v^{\prime}-1,1/2}\rangle$ and $|g^{-}_{v^{\prime},3/2}\rangle \longleftrightarrow |e^{+}_{v^{\prime}-1,1/2}\rangle$ transitions to pump population back into the cycling transition, as shown by the R$1$, R$2$ and unlabeled arrows in figure \ref{fig:coolingoneparity}.

However, when decay between vibrational levels is fast, the resulting parity flips may populate the $|g^{+}_{v^{\prime},J'}\rangle$ states, so lasers connecting these states to an excited state must also be included to avoid population buildup in the even-parity states.  Multi-stage vibrational decay also results in angular-momentum diffusion.  For instance, decay from the $|e^{+}_{0,1/2}\rangle$ populates the $|g^{-}_{1,3/2}\rangle$ state, which decays into the $|g^{+}_{0,5/2}\rangle$ state.  Population is pumped out of states with $J'>3/2$ by driving $|g^{\pm}_{0,J'}\rangle \longleftrightarrow |e^{\mp}_{0,J'-1}\rangle$.   This cooling scheme is illustrated in figure \ref{fig:coolingtwoparity}.

\begin{figure}
\center
\includegraphics[width=0.65\textwidth]{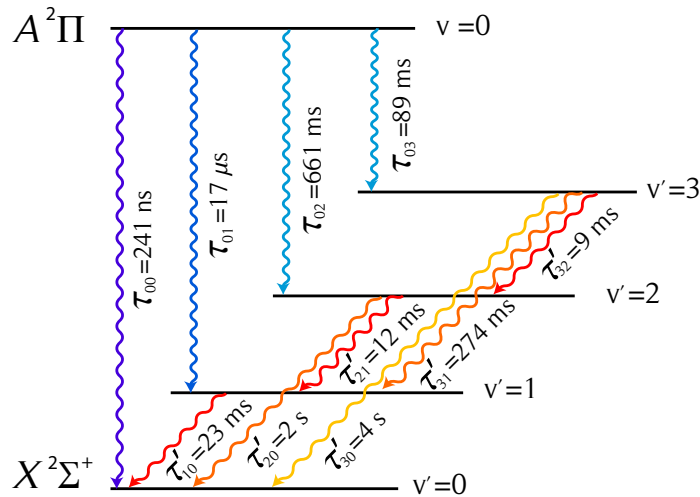}
\caption{Energy level schematic showing rotationless lifetimes of different BH$^+$ levels (not to scale).  Lifetimes for A$\rightarrow$X transitions are denoted by $\tau$ and lifetimes for transitions within the ground state are denoted by $\tau '$. All lifetimes are calculated using the \emph{ab initio} potential energy curves and dipole moments discussed in section~\ref{subsec:methods}. \label{fig:BHlifetimes}}
\end{figure}

\begin{figure}
\center
\includegraphics[width=0.65\textwidth]{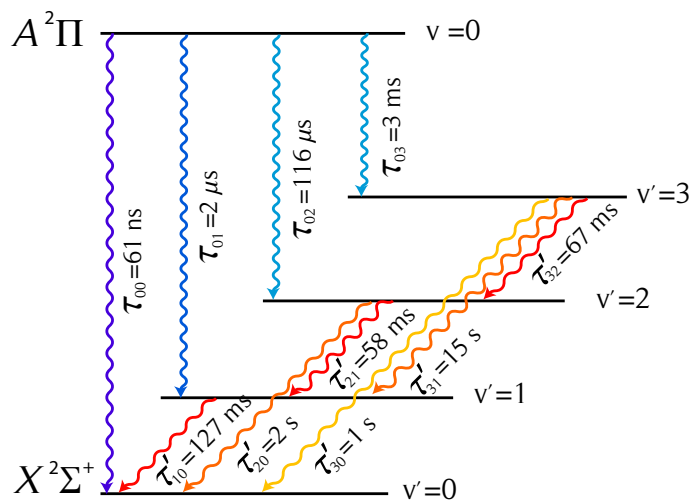}
\caption{Energy level schematic showing rotationless lifetimes of different AlH$^+$ levels (not to scale).  Lifetimes for X$\rightarrow$A transitions are denoted by $\tau$, and lifetimes for transitions within the ground state are denoted by $\tau '$.  All lifetimes are calculated using the \emph{ab initio} potential energy curves and dipole moments discussed in section \ref{subsec:methods}. \label{fig:AlHlifetimes}}
\end{figure}

In figures \ref{fig:BHlifetimes} and \ref{fig:AlHlifetimes} we show the rotationless lifetimes for \bhp\ and \alhp\, respectively. Since both \alhp\ and \bhp\ have relatively fast decay rates (up to $100\ \mathrm{s}^{-1}$) between vibrational levels, we find that they require cycling and repumping on both parities. In the following section, we examine the details of each molecule separately, and determine their repumping requirements. Predissociation and photodissociation rates are considered in the last subsection.

\subsection{\label{subsec:sim-results}Simulation results}
\subsubsection{\bhp}

\begin{figure}
\center
\includegraphics[width=0.65\textwidth]{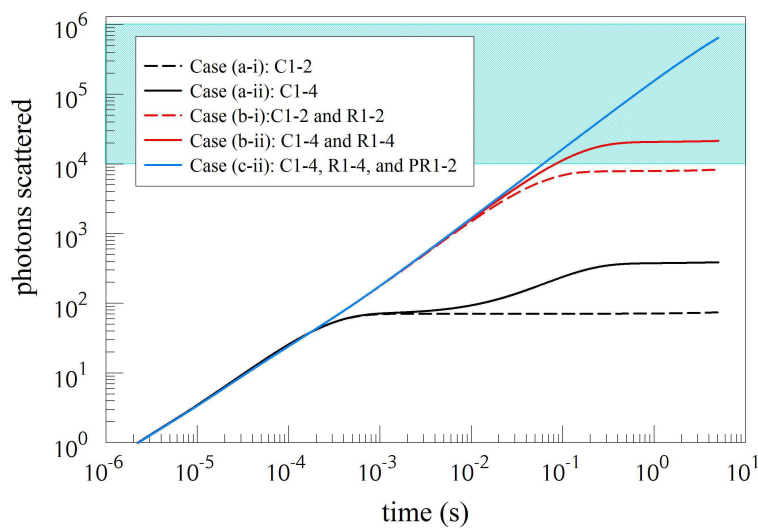}
\caption{ Plot of the integrated cycling-transition photon count for the \bhp\ simulation using different laser configurations, driving subsets of transitions shown in figure \ref{fig:coolingtwoparity}.  The hatched region corresponds to upper- and lower-limits for Doppler cooling to mK temperatures, corresponding to fixed and swept detuning, respectively. \label{fig:bhsimulation}}
\end{figure}

In figure~\ref{fig:bhsimulation} the total number of scattered photons for different laser configurations is plotted.  Population is initially in the $|g^{-}_{0,1/2}\rangle$ state.  The simulation includes dipole-allowed transitions between electronic states, and vibrational and rotational transitions within the X$^{2}\Sigma^{+}$ state.  Blackbody radiation is also included in the model, for a temperature of $300\ \mathrm{K}$.  In Case (a-i), two wavelengths are used to drive the odd-parity cycling transitions: C$1$ and C$2$ in figures~\ref{fig:coolingoneparity} and~\ref{fig:coolingtwoparity}.  Population is quickly pumped into the $v^{\prime}=1$ manifold, and a total of $N\approx 70$ photons are scattered.  In Case (a-ii), an additional two wavelengths are introduced to drive the even-parity cycling transitions: C$3$ and C$4$ (see figure \ref{fig:coolingtwoparity}).  Again, population is quickly pumped into the $v^{\prime}=1$ manifold; however, at longer times ($t \approx \tau^{\prime}_{10}$), population leaks back into the even-parity ground states and is driven by the even-parity cycling transitions, resulting in a total of $N\approx 380$ scattered photons.  The delay between successive photon count plateaus is due to population building up in the $v^{\prime}=1$ manifold.

Case (b-i) includes the transitions from Case (a-i) plus an additional two wavelengths to drive the odd-parity repumping transitions: R$1$ and R$2$ in figures~\ref{fig:coolingoneparity} and~\ref{fig:coolingtwoparity}. $N\approx 8000$ photons are scattered, with the dominant dark-state leak being the even-parity ground states.    Case (b-ii) extends Case (a-ii) with an additional four wavelengths to drive even- and odd-parity repumping transitions: R$1$-R$4$.  A total of $N\approx 21000$ photons are scattered, which plateaus at just above the swept-detuning threshold.  

As predicted in section~\ref{subsec:cooling-scheme}, vibrational decay plays a significant role in depopulating the higher-lying vibrational ground states, resulting in parity flips and rotational diffusion.  In Case (b-ii), the parity-flip issue is addressed, so population built up in the $|g_{0,5/2}^{\pm}\rangle$ states.  Case (c-ii) includes the same wavelengths as Case (b-ii) and also two additional wavelengths to drive odd- and even-parity P-branch repumps: PR$1$, PR$2$ (see figure \ref{fig:coolingtwoparity}).  The increase in counts $(N\approx 600,000)$, which is well above the swept-detuning threshold required for significant cooling from room temperature.

\subsubsection{\alhp}

\begin{figure}
\center
\includegraphics[width=0.65\textwidth]{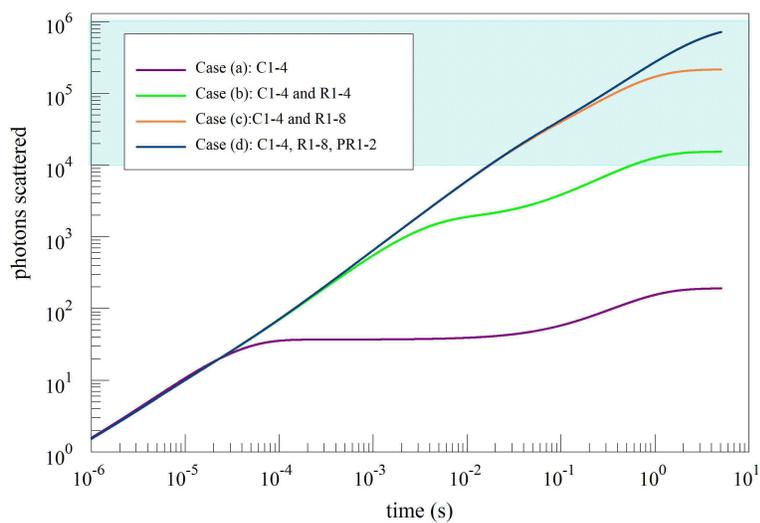}
\caption{Plot of the integrated photon count for the \alhp\ simulation using different laser configurations, driving subsets of transitions shown in figure \ref{fig:coolingtwoparity}.  The hatched region corresponds to upper- and lower-limits for Doppler cooling to mK temperatures, corresponding to fixed and swept detuning, respectively. \label{fig:alhsimulation}}
\end{figure}

The simulation of \alhp\ is plotted in figure \ref{fig:alhsimulation}.  Again, we assume that population starts in the $|g^{-}_{0,1/2}\rangle$ state.  The simulations include transitions between electronic states, within the ground electronic state, and include a blackbody source of radiation at $300\ \mathrm{K}$.  In Case (a) four wavelengths are used to drive only the cycling transition: C$1$-C$4$.  A total of $N\approx 200$ photons are scattered.

In Case (b) an additional four wavelengths are used to drive the repumping transitions: R$1$-R$4$.  There is a significant increase in counts ($N\approx 15000$), and a noticeable deviation from the results of \bhp. At $t\approx 10^{-3} \ \mathrm{s}$, there is a reduction in rate, resulting from population building up in the $v'=2$ level. For BH$^+$ the rate of populating $v^\prime=2$ is $55$ times \emph{slower} than the decay from $v^\prime=2$ to $v^\prime=1$ ($\tau_{02}/\tau^\prime_{21}=55$). For AlH$^+$ the rate of populating $v^\prime=2$ is $500$ times \emph{faster} than the decay from $v^\prime=2$ to $v^\prime=1$ ($\tau_{02}/\tau^\prime_{21}=1/500$), resulting in population trapping.

In Case (c) an additional four wavelengths are applied to drive the repumping transitions: R$5$-R$8$.  This results in a total of $N\approx 210,000$ photon counts, which is well above the swept-detuning cooling threshold.  Interestingly, the second vibrational repump is sufficient for cooling of \alhp\, whereas \bhp\ requires optical pumping of the $|g^{\pm}_{5/2}\rangle$ states.

In Case (d) $\Delta J = -1$ repumping transitions are included: PR$1$ and PR$2$.  The resultant photon count ($N\approx 700,000$), which is similar to the number of counts we obtain in Case (c-ii) of \bhp\ .  We find that the results of \alhp\ are similar to those of \bhp\ in that they each require repumping of both parities, due to vibrational decay; however, \alhp\ differs in that it requires an additional vibrational repump, while not requiring additional rotational repumping.

\subsubsection{\label{subsec:disastrous-events}Calculation of predissociation and photofragmentation rates}
The cooling scheme requires a continuous repopulation of the vibrational ground level of the \AdP\ excited electronic state. One possible mechanism that stops the cooling cycle is isoenergetic predissociation of the molecular ion by the coupling of the bound excited state to a dissociative state of the \XdS\ repulsive wall. Bound and continuum wave functions can interact via spin- and rotational-orbit couplings. The expression for such predissociation rates, in units of s$^{-1}$, is obtained from Fermi's golden rule ~\cite{bcont2.2}

\begin{eqnarray}
k_s(v,J)=\frac{2\pi}{\hbar}|\langle\psi_{E,J'}(R)|M_s(R)|\psi_{v,J}(R)\rangle|^2,
\label{eq:prediss}
\end{eqnarray}

\noindent where $M_s(R)$ is a term in the molecular Hamiltonian, which is neglected in the Born-Oppenheimer approximation and  describes the coupling between the initial and final electronic states. The $|\psi_{v,J}(R)\rangle$ initial-state wavefunction is space-normalized and the $|\psi_{E,J'}(R)\rangle$ wave function of the continuum is energy-normalized~\cite{Field's_book}. The calculation uses a one-dimensional density of states in the bond coordinate to determine the energy-normalization, $d(E)\propto \sqrt{\mu/(E-D)}$, where $\mu$ is the reduced mass and $D$ is the dissociation energy of the final electronic state \cite{bcont2.2}.

The predissociation depends strongly on the slope of the repulsive part of the dissociative potential. In order to estimate the uncertainty in the predissociation calculations, we first calculate the overlap integral of the wavefunctions assuming and multiplying by a bond-length independent spin-orbit coupling of 13.9 cm$^{-1}$~\cite{Ramsay1982}. For \bhp, by varying the details of the bound-state wavefunction, we find that the predissociation rate varies by more than four orders of magnitude from 10$^2$ to 10$^{-2}$ s$^{-1}$.  Representing the \AdP(v=0) wave function by a symmetric Gaussian from the solution of the harmonic oscillator, the rate results in the most disastrous scenario, $4.4\times10^2$ s$^{-1}$. Using the solution of a Morse potential gives $6.2\times 10^{-2}$ s$^{-1}$. Combining the continuum and bound wavefunctions computed using LEVEL 8.0 and BCONT 2.2, respectively, and by inputting the FCI(3e$^-$)/aug-cc-pV5Z \AdP\ and \XdS\ PECs, the predissociation rate is $1.6\times 10^{-1}$ s$^{-1}$ (potentials and wavefunctions shown in figure~\ref{fig:fciBH}). Including the spin-orbit coupling function calculated in section~\ref{subsec:abinitio}, as prescribed in equation (\ref{eq:prediss}), the predissociation rate results in a value of 0.03$\pm$0.02 s$^{-1}$. For \alhp\ (using EOM-CC3 potential energy functions and MRCI spin-orbit coupling matrix elements), the predissociation rate is 0.2$\pm$0.1 s$^{-1}$. The quoted uncertainties come from interpolation noise in both the repulsive part of the \XdS\ state and the point piecewise polynomial used to fit the R-parametrized spin-orbit couplings. This uncertainty due to interpolation is negligible for the calculation of optical decay rates ($\sim$10 parts per million) reported elsewhere in this manuscript.

\begin{figure}
\centering
\includegraphics[width=0.65\textwidth]{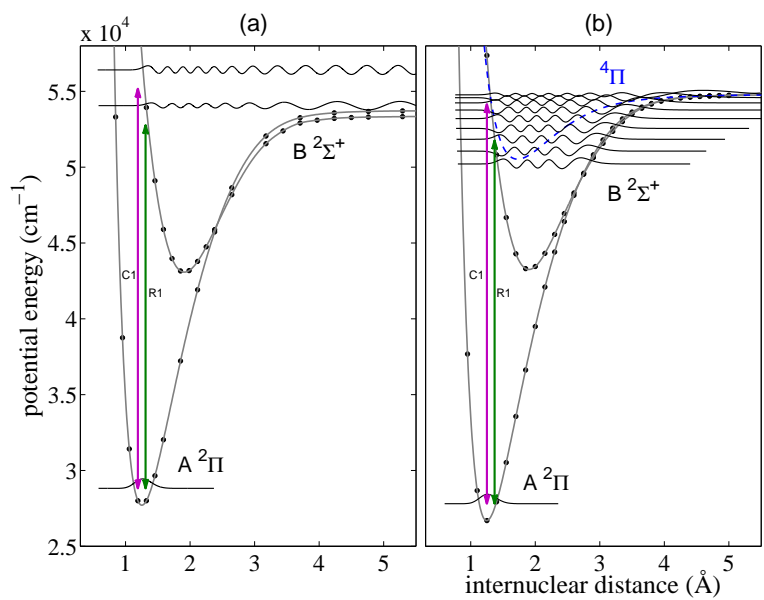}
\caption{\label{fig:secondphotdiss} Sensitivity of problematic two-photon channels to precise details of molecular structure.  Potential energy functions of the $\Lambda=1$ electronic states of \bhp\ are plotted at the a) MC-SCF level from \cite{KleinJCP1982}, and b) FCI(3e$^-$)/aug-cc-pV5Z. Black dots are the actual \abinitio\ results and the solid lines are splines with analytical functions described in the BCONT 2.2 and LEVEL 8.0 manuals~\cite{bcont2.2,level8.0}. The wavefunctions shown, which are used to calculate photofragmentation or photoabsorption rates, are positioned at the eigenvalues that solve the radial Schr\"{o}dinger equation. The length of the arrows represents the actual experimental energy of the C$1$ and R$1$ photons.}
\end{figure}

Another problematic event is the excitation from the \AdP\ state to either bound or dissociative \BdS\ levels, by absorption of a cooling or repump photon. For \bhp, PECs from two different \abinitio\ levels of theory place the dissociation limit, V$_{lim}$, either below or above the energy accessible by the 379 and 417 nm photons exciting from the A state (see table~\ref{tab:secphotdiss} and figure~\ref{fig:secondphotdiss}). Although both results agree with predictions based on experimental observations (due to a 2000 \cm\ uncertainty), these calculations yield two different photon absorption scenarios. The first (MC-SCF) would photodissociate the molecule, and the second (FCI) would steal population from the cooling cycle to a bound vibrationally excited level of the \BdS\ state. For \alhp, MC-SCF and EOM-CC3 predict V$_{lim}$ to be below the total energy accessed by any of the three required cooling and repump wavelengths (360, 381 and 376 nm).

\begin{table}
\caption{\label{tab:secphotdiss} Energies and rate coefficients relevant to possible photofragmentation events. $\hbar\omega$ (T), $A$ (T), and $I_{s}$ (T) correspond to the photon energy, $\mathcal{A}$ Einstein coefficient, and saturation intensity for each \APiXSig\ transition T as labeled in figures ~\ref{fig:coolingoneparity} and \ref{fig:coolingtwoparity}. The \bhp\ cooling schemes do not involve the transition R$5$. M$^+$($^3$P)-M$^+$($^1$S) is the energy difference between the lowest singlet and triplet energy states of  B$^+$ or Al$^+$. This should correspond to the energy difference between the dissociative asymptotes of the \XdS\ ground state and the common \AdP\ and \BdS\ excited states V$_{lim}$. $E_{total}$ is the energy of the \AdP\ ($v=0,J=1/2$) state plus the photon energy of the transition T. Photofragmentation does not occur in the FCI(3e$^-$)/aug-cc-pV5Z) calculation for \bhp\ (see text), but the \XdS\ D$_e$ does not preclude either possibility due to current experimental uncertainties.}
\tiny

\begin{tabular}{@{}lllllll}
 & \multicolumn{3}{c}{\bhp} & \multicolumn{3}{c}{\alhp} \\
\cline{2-4}
\cline{5-7}\\
 & MC-SCF$^{\rm{a}}$& FCI & Observed$^{\rm{b}}$& MC-SCF$^{\rm{a}}$ & EOM-CC3/aug-cc-pCVQZ & Observed$^{\rm{c}}$ \\
\mr
\XdS(v=0,K=1) [\cm] & 1223 & 1268 & 1272 & 810 & 848 \\
\XdS(v=2,K=1) [\cm] &  &  &  & 3669 & 3702 \\
$\hbar\omega$ (C1) [\cm] & 27594 & 26547 & 26356 & 28521 & 27747 & 27673 \\
$\hbar\omega$ (R1) [\cm] & 25239 & 24163 & 23960 & 27001 & 26225 \\
$\hbar\omega$ (R5) [\cm] &  &  &  & 27255 & 26566 \\
$\mathcal{A}$ (C1) [s$^{-1}$]  & 2.9$\times10^6$ & 2.8$\times10^6$ &  & 1.1$\times10^7$ & 1.1$\times10^7$ \\
$\mathcal{A}$ (R1) [s$^{-1}$]  & 3.8$\times10^4$ & 3.9$\times10^4$ &  & 1.3$\times10^5$ & 2.9$\times10^5$ \\
$\mathcal{A}$ (R5) [s$^{-1}$] &  &  &  & 4.8$\times10^5$ & 8.7$\times10^5$ \\
$I_s$ (C1) [photons/(cm$^2$s)] & 2.3$\times10^{15}$ & 2.0$\times10^{15}$ &  & 9.6$\times10^{15}$ & 8.8$\times10^{15}$ \\
$I_s$ (R1) [photons/(cm$^2$s)] & 2.6$\times10^{13}$ & 2.4$\times10^{13}$ &  & 1.0$\times10^{14}$ & 2.1$\times10^{14}$ \\
$I_s$ (R5) [photons/(cm$^2$s)]&  &  &  & 3.7$\times10^{14}$ & 6.4$\times10^{14}$ \\
\XdS\ D$_e$ [\cm] & 15645 & 17261 & 17000$\pm$2000 & 5998 & 6694 & $\approx$6300 \\
M$^+$($^3$P)-M$^+$($^1$S) [\cm] & 38068 & 37546 & 37342 & 37003 & 37206 & 37454 \\
A and B V$_{lim}$ [\cm] & 53713 & 54807 & 54450$\pm$2000 & 43000 & 43900 & $\approx$43750 \\
 E$_{Total}$ (C1) [\cm] & 56411 & 54362 & 53984 & 57852 & 56342 \\
 E$_{Total}$ (R1) [\cm] & 54056 & 51978 & 51588 & 56332 & 54820 \\
 E$_{Total}$ (R5) [\cm] &  &  &  & 58179 & 56834 \\
$\sigma_{abs}$ (C1) [cm$^2$]  & 1.1$\times10^{-20}$ &  &  & 7.2$\times10^{-23}$ & 3.5$\times10^{-22}$ \\
$\sigma_{abs}$ (R1) [cm$^2$]  & 7.5$\times10^{-21}$ &  &  & 1.4$\times10^{-22}$ & 7.1$\times10^{-22}$ \\
$\sigma_{abs}$ (R5) [cm$^2$]  &  &  &  & 5.4$\times10^{-22}$ & 2.4$\times10^{-21}$ \\
$k_{diss}$ (C1) [s$^{-1}$]  & 2.6$\times10^{-5}$ &  &  & 6.9$\times10^{-7}$ & 3.1$\times10^{-6}$ \\
$k_{diss}$ (R1) [s$^{-1}$] & 1.9$\times10^{-7}$ &  &  & 1.4$\times10^{-8}$ & 1.5$\times10^{-7}$ \\
$k_{diss}$ (R5) [s$^{-1}$] &  &  &  & 2.0$\times10^{-7}$ & 1.6$\times10^{-6}$ \\
\end{tabular}

\noindent $^{\rm{a}}$ \cite{KleinJCP1982} \\
\noindent $^{\rm{b}}$ \cite{RosmusJCP1977,Ramsay1982,ViteriJCP2006} \\
\noindent $^{\rm{c}}$ \cite{AlmyPR1934,RosmusJCP1977}

\end{table}

The photodissociation cross section (in cm$^2$/molecule), is~\cite{LeRoyJCP1976}:

\begin{eqnarray}
\nonumber
\sigma_s(v,J)=\frac{2 \pi \omega}{3\hbar c}\sum_{J'}\frac{S_J^{J'}}{2J+1} |\langle\psi_{E,J'}(R)|M_s(R)|\psi_{v,J}(R)\rangle|^2 \\
%\sigma_s(v,J)&=&3.22696\times10^{-20}\nu\sqrt {\mu/K_s}\frac{S_J^{J'}}{2J+1} \\
% & & \times \left|\int_0^\infty \! \psi_{E,J'}^*(R)M_s(R)\psi_{v,J}(R)\, \mathrm{d}R\right|^2,
 \label{eq:photodiss}
\end{eqnarray}

\noindent $S$ is the usual H\"onl-London rotational intensity factor ($S_J^{J'}/(2J+1)$ is set to 1 in this section), and $M_s(R)$ is the transition moment function,  including the ratio of initial to final-state electronic degeneracy factors. Allowed values of $J'$ are given by the usual rotational selection rule. Similar to equation (\ref{eq:prediss}), this expression assumes that the continuum radial wavefunction amplitude is energy normalized.

Assuming saturation intensities, calculated from Einstein \A\ coefficients for the main cycling and repump transitions, we calculate the photon flux for each transition, $I_s$ (T), listed in table~\ref{tab:secphotdiss}. The photodissociation rates are the multiplication of these laser intensities by the cross sections calculated from equation (\ref{eq:photodiss}). Continuum and bound wavefunctions were calculated using BCONT 2.2 and LEVEL 8.0, respectively, and using the EOM-CC3 potential energy curves and the MRCI transition moment functions. We perform simulations using the MC-SCF potential energies and transition dipole moments from \cite{KleinJCP1982} for comparison.  We determine that photodissociation occurs on a timescale of tens of hours for \bhp, and several days for \alhp.  

For \bhp, the FCI potential energy curves predict that the cycling and repump wavelengths result in an excitation from the \AdP\ state into a bound level of the \BdS\ state.  We believe that the probability of resonant absorption to the \BdS state is small due to the narrow laser linewidths relative to the vibrational energy spacing and due to the poor Franck-Condon overlap. The most intense $B-A$ band reaching this energy region of the \BdS\ state is the $(11,0)$ with the value for \A\ of $1.6\times10^3$ s$^{-1}$.   Population of this state results in vibrational diffusion which is not repumped in
our cycling scheme.

Given the 2000 \cm\ experimental uncertainty in the dissociation energy, we do not have enough information to determine if the photoabsorption (predicted by FCI) or photodissociation (according to MC-SCF) processes will present a problem for Doppler cooling

\section{Prospects for Cooling}
\label{sec:prospects}
Our simulations show that we require a total of eight wavelengths to Doppler cool \bhp\ and \alhp (see figures ~\ref{fig:bhsimulation} and~\ref{fig:alhsimulation}); however, cooling does not necessarily require eight unique lasers.  For both ions, two of the cycling transitions, C$1$ and C$2$, can be addressed by a single laser, since these two wavelengths differ only by the spin-rotation splitting ($\Delta \omega_{sr}\approx 540\ \mathrm{MHz}$ for \bhp~\cite{Ramsay1982} and  $\Delta \omega_{sr}\approx 1.7\ \mathrm{GHz}$ for \alhp~\cite{AlmyPR1934}). This splitting can be achieved from a single laser source by using either an acousto-optic modulator (AOM), or an electro-optic modulator (EOM).  The other two cycling transitions, C$3$ and C$4$ each require their own source since their splitting is large ($\Delta \omega \approx 1\ \mathrm{THz}$ for \bhp\ and $\Delta \omega \approx 2\ \mathrm{THz}$ for \alhp). The cycling transitions require CW lasers that are narrow relative to the electronic linewidth.

\bhp\ requires driving the transitions R$1$-$4$, corresponding to repumping $v^{\prime}=1$. These transitions also span 1 THz, so they would require a total of three CW lasers, since spin-rotation splitting between R$2$ and R$3$ can be bridged by an AOM.  However, repumping on all four transitions can be achieved with a single femtosecond pulsed laser, with a typical bandwidth of $100\ \mathrm{cm^{-1}} (3\ \mathrm{THz})$. Using femtosecond repumps generally requires optical pulse shaping, in order to avoid R-branch (rotational heating) transitions.  Pulse-shaped pumping has been used to achieve vibrational cooling of Cs$_{2}$ molecules, with a resolution of $1\ \mathrm{cm}^{-1}\ (30\ \mathrm{GHz})$ \cite{Science.321.232} -- which is sufficient resolution for rotationally-resolved repumping of hydrides. A femtosecond laser can also be used to drive the P-branch repumping transitions PR$1$ and PR$2$; however, these transitions would require a second laser since the R$1$-$4$ transitions differ from the PR$1$ and PR$2$ by more than $100\ \mathrm{cm}^{-1} \ (3\ \mathrm{THz})$.  It should noted that repumping of PR$1$ and PR$2$ may not be necessary, since the total photon count exceeds the number required for Doppler cooling in the swept-detuning case (see figure~\ref{fig:bhsimulation}).

\alhp\ requires driving transitions R$1$-$4$, corresponding to repumping $v^{\prime}=1$ and R$5$-$8$, corresponding to repumping $v^{\prime}=2$. The repump R$1$-$4$ span $2\ \mathrm{THz}$, which is within the bandwidth of a femtosecond laser.  Transitions R$5$-$8$ also span $2\ \mathrm{THz}$; however they do not overlap with transitions R$1$-$4$ ($\Delta \omega \approx 9\ \mathrm{THz}$), so they would require a second femtosecond laser.

If a broadband source for repumping is used in conjunction with CW lasers for cooling, the laser requirements for \bhp\ and \alhp\ are reduced to five and six lasers driving 10 and 14 different transitions, respectively. For \bhp, repumping transitions R$1$-R$4$ require one femtosecond laser, and repumping transitions PR$1$-$2$ require a second femtosecond laser. For \alhp, repumping transitions R$1$-R$4$ require one femtosecond laser, and repumping transitions R$5$-R$8$ require a second femtosecond laser. The PR$1$-$2$ wavelengths would require a third femtosecond laser but they are not strictly necessary (see figure~\ref{fig:alhsimulation}).

The comb-like nature of the frequency components of a femtosecond laser can be problematic if a transition which we wish to drive happens to fall between two comb lines (typically separated by $80\ \mathrm{MHz}$).  Various modulation techniques, either internal or external to the femtosecond laser, can be used to address this complication.

\section{Conclusions}
\label{sec:conclusions}
The key element for direct laser cooling of molecules~\cite{DiRosaEPJD2004, StuhlPRL2008, ShumanPRL2009, ShumanNAT2010} and molecular ions are diagonal FCFs. From our survey of a number of molecules (see \ref{app:pot-candidates}) we determined  \bhp\ and \alhp\ to be favorable candidates since a probabilistic FCF analysis (excluding vibrational decay) suggested that three and two vibrational bands, respectively, needed to be addressed to scatter $10^4$ photons.  A detailed analysis of state population dynamics (including vibrational decay) has shown that \bhp\ requires addressing two vibrational bands while \alhp\ requires addressing three.  For both molecular ions we find that vibrational decays result in diffusion of both parity and angular momentum; thus, implementation of this cooling scheme requires addressing the molecules with a large number of repump wavelengths. 

To minimize the total number of lasers required, a femtosecond laser can be used, since multiple transitions fall within the laser bandwidth.  However, for many of these transitions, the frequency of excitation needs to first be experimentally determined with higher accuracy.  The spectroscopy can be performed and the direct cooling technique optimized by initially trapping the molecular ions with laser-cooled atomic ions and monitoring the sympathetic heating and cooling of the atomic ion by the molecular ion ~\cite{ClarkPRA2010}.  Ultimately, dissociation of the molecular ion represents a fundamental limit for continuous Doppler cooling of the species presented here.

The attempted laser-cooling of BH$^+$ and AlH$^+$ will either result in the first directly laser-cooled molecular ions and/or and accurate measurement of rare predissociation events. Either way it will improve our knowledge of the spectroscopic splittings of BH$^+$ and AlH$^+$ by an order of magnitude. Together with experimental dissociation rates, measurements of these splittings yield the most stringent test of advanced electronic structure theories; as the five-valence-electron open-shell diatomic \bhp\ is among the simplest stable molecules and serves as a prototype for the development and testing of such theories~\cite{RosmusJCP1977, GuestCPL1981, KleinJCP1982, KusunokiCPL1984, CooperCPL1986, CurtissJCP1988, FellerJCP2000, HirataCPL2000, IshidaCPL2001, PetsalakisMP2006}. 

More generally, any molecular ion can be sympathetically cooled, and the long trapping lifetime allows slow processes to be investigated.  Thus, sympathetically cooled molecular ions could serve as a valuable tool in determining the rates of higher-order processes such as predissociation and photofragmentation, which are of concern for direct Doppler cooling applications.

\ack
We would like to thank Eric Hudson for bringing to our attention the potential problem of predissociation, and for helping us with initial calculations of predissociation rates. KRB and CRV were supported by Georgia Tech and the NSF (CHE-1037992).  JHVN and BO were supported by the David and Lucile Packard Foundation (Grant No. 2009-34713) and the AFOSR (Grant No. FA9550-10-1-0221).

\appendix
\section{\label{app:pot-candidates}Potential Candidates}
Doppler cooling candidates are divided into two main categories. The first category corresponds to transitions in which the electron is excited from the HOMO to the LUMO, and the second category corresponds to transitions in which a hole moves from one orbital to another. Each category is subdivided according to grouping on the periodic table. Groups which we discuss are chosen, based on chemical intuition, to form open-shells with vertical excitations that do not perturb the chemical bond. Probabilistic predictions in this appendix are based on FCFs obtained from Morse potentials defined by spectroscopic constants available in the literature. Detailed population-dynamics studies are needed for a more rigorous selection.

\subsection{Diatomic molecular ions with transitions which excite a single electron from the HOMO to the LUMO}
\subsubsection{Group 3 hydrides and halides}
ScH$^+$ has an interesting \dD-\dP\ system with very similar spectroscopic constants for both electronic states; however, the transition is deep into the IR ($\lambda=6~\mu m$)~\cite{Alvarado-SwaisgoodJPC1985}. The ground state of YH$^+$ is a \dS\ and the first excited state is a \dD. Spectroscopic constants \re\ and \we\ are similar but the transitions falls in the IR at $\approx 3.3~\mu m$~\cite{PetterssonJCP1987}. Finally, LaH$^+$ has a \dSp-\dD\ system with similar spectroscopic constants and the transition is at $\approx 5.1~\mu m$~\cite{DasJCP1991}.

\subsubsection{Group 13 hydrides and halides}
The open-shells from Group 13 and halogens are stable with \dS\ ground states. Unfortunately, the \dP\ potential curve is dissociative. This is common in the following examples: AlCl$^+$, AlF$^+$~\cite{GlenewinkelmeyerJCP1988, GlenewinkelmeyerJCP1991} and GaCl$^+$. Although GaF$^+$ is believed to have strong ionic character~\cite{MochizukiTCA1999} it also has a dissociative first exited state~\cite{YoshikawaCPL1995}. The hydrides \bhp\ and \alhp\ seem to be suitable for direct laser cooling and are the subject of the present paper. Experimental spectroscopic data  for GaH$^+$ is not yet available~\cite{MochizukiTCA1998}.

\subsubsection{Bialkalis, alkali monohalides, alkali monohydrides and molecular ions formed by alkalis and Group 13}
Li$^+_2$ and LiNa$^+$ have \dSpg\ and \dSp\ ground electronic states with only one valence electron, and they do not have valence correlation energy~\cite{BoldyrevJCP1993}. Their ground state dissociation energies are relatively high (~10000 \cm) and they do not seem to have dissociative states that couple non-adiabatically to the \APiXSig\ system. The lack of spectroscopic data on these systems stop us from making further predictions, but it would not be surprising if a few of these molecular ions have diagonal FCFs. These systems can not be ruled out as potential candidates. The same seems to be true for LiAl$^+$ with a bound ground state and a dissociation energy of roughly 1 eV~\cite{BoldyrevJCP1993}. LiB$^+$ has a weaker bond and the A and X states spectroscopic constants are different~\cite{CaoIJQC1998}.

With respect to the alkali monohydrides, although they are technically open-shells with one unpaired electron spin, they have very weak bonds. This is expected as the same unpaired electron shares some density to participate in the bond (e.g., KH$^+$~\cite{KorekJMST2008} and NaH$^+$~\cite{MagnierJPCA2005})

\subsubsection{Group 15 halides and hydrides}
PH$^+$ and PF$^+$ would require repumping of six and ten vibrational bands to scatter 2$\times 10^4$ and 9$\times 10^4$ photons, respectively. They do not have nuclear spin and their ground states are $^2\Pi$. The first excited state of PH$^+$ is a $^2\Delta$ and the vertical transition from the ground state is at 381 nm. For the PF$^+$ case, the A$^2\Sigma$-X$^2\Pi$ optical transition is in the UV at 282 nm~\cite{Radzigs_book}.

\subsection{Diatomic molecular ions with an unpaired electron in which an optical transition moves a hole to a bonding orbital}
\subsubsection{Group 17 hydrides}
Using Morse potentials derived from spectroscopic constants compiled by Radzigs and Smirnov~\cite{Radzigs_book}, we notice A$^2\Sigma$-X$^2\Pi$ transitions in molecules like HF$^+$, HCl$^+$ and HBr$^+$ to have very un-diagonal FCFs. However, a recent theoretical proposal to measure molecular parity violation~\cite{IsaevPRA2010} points out that HI$^+$ could have highly diagonal FCFs. This is attributed to the center of mass being very close to the iodine atom, resulting in the molecular orbit having mostly an atomic character. There is not enough spectroscopic information of excited states of HI$^+$ to design a laser cooling experiment~\cite{ChandaJCP1995}.

\subsubsection{Group 15 diatomic ions}
We use FCFs from spectroscopic constants compiled by Radzigs and Smirnov~\cite{Radzigs_book} to test N$^+_2$ and P$^+_2$. The former could scatter 11$\times 10^4$ photons if irradiated by five lasers. These transitions fall in the near IR with the most energetic transition at 1091 nm. This homonuclear molecular ion has recently been loaded and sympathetically cooled in its rovibrational ground state~\cite{TongPRL2010}. Although P$^+_2$ has a $^2\Pi_u$ ground state with similar \re\ and \we\ constants as its first \dSpg\ excited state, the FCFs show that one would have to address more than 8 transitions with $\lambda > 4.6 \mu m$ to scatter over $10^4$ photons.

\subsubsection{Group 5 and 6 halides}
These molecules have crowded excited state manifolds and the first electronic states high multiplicity. VF$^+$ has a $A ^4\Delta-X ^4\Pi$ system with very similar \re\ and \we\ constants and a vertical transition at $\approx 7~\mu m$. If it is possible to cope with higher multiplicity (S=5), CrF$^+$ might work as it has a $A ^5\Pi-X ^5\Sigma^+$ with an IR transition ($\lambda=2~\mu m$)~\cite{KardahakisJCP2005}.

\bibliographystyle{unsrt}
\bibliography{ion-laser-cool}
\end{document}